\documentclass[conference,letter]{IEEEtran}
\usepackage{bm}
\usepackage{amsmath, amssymb}
\usepackage{bm}
\usepackage{upgreek}
\usepackage{amssymb}
\usepackage{xcolor}
\usepackage{pgfplots} \pgfplotsset{compat=newest}
\usepackage{tikz}
\usetikzlibrary{plotmarks} \usetikzlibrary{arrows.meta}
\usepgfplotslibrary{patchplots} 
\usepackage{grffile}
\usepackage{epstopdf}
\usepackage{comment}
\newcommand{\vect}[1]{\boldsymbol{#1}}
\newcommand{\norm}[1]{\left\lVert#1\right\rVert}

\usepackage{cite}

\ifCLASSINFOpdf
   
\else
\fi

\begin{document}
%
\title{RIS-Enabled Localization Continuity\\ Under Near-Field Conditions}

\author{\IEEEauthorblockN{Moustafa Rahal\IEEEauthorrefmark{1}\IEEEauthorrefmark{3}, Beno\^{i}t Denis\IEEEauthorrefmark{1}, Kamran Keykhosravi\IEEEauthorrefmark{2}, Bernard Uguen\IEEEauthorrefmark{3}, Henk Wymeersch\IEEEauthorrefmark{2}}
\IEEEauthorblockA{\IEEEauthorrefmark{1} 
CEA-Leti, Université Grenoble Alpes, F-38000 Grenoble, France\\
\IEEEauthorrefmark{2} 
Department of Electrical Engineering, Chalmers University of Technology, Gothenburg, Sweden\\
\IEEEauthorrefmark{3} 
Université Rennes 1, IETR - UMR 6164, F-35000 Rennes, France\\
}

}
\maketitle

\begin{abstract}
Reconfigurable intelligent surfaces (RISs) have the potential to enable user localization in scenarios where traditional approaches fail. Building on prior work in single-antenna RIS-enabled localization, we investigate the potential to exploit wavefront curvature in geometric near-field conditions. Via a Fisher information analysis, we demonstrate that while near-field improves localization accuracy \textcolor{black}{mostly at short distances when the line-of-sight (LoS) path is present, it could still provide reasonable performance when this path is blocked by relying on a single RIS reflection.}
After deriving and illustrating the corresponding position error bounds as a function of key operating parameters, we discuss practical system approaches that could enable better LoS-to-NLoS positioning continuity in harsh environments.
\end{abstract}

%
\IEEEpeerreviewmaketitle

\section{Introduction}

Reconfigurable intelligent surfaces (RISs), which consist of (semi-)passive low-complexity components such as reflect-arrays or transmit-arrays (e.g., typically, with inter-element spacing equal or lower than half the wavelength of transmitted signals), can be used to purposefully adjust radio propagation channels \cite{2019Basar}. RISs can behave as controllable electromagnetic mirrors to create anomalous reflections, as refracting lenses to limit the number and complexity of radio-frequency chains in reception, or even as transmittive (though non-regenerative) relays. 
Accordingly, RISs have been identified as a flexible breakthrough technology capable of shaping sustainable radio environments, as a new kind of service provisioning in future 6G communication networks~\cite{RISE-6G-2021}. However, many questions are still outstanding, e.g., in terms of models and use cases~\cite{bjornson2020reconfigurable}. 

The main motivation for using RIS so far has been to improve communication-relevant metrics, in particular when the line-of-sight (LoS) path between the base station (BS) and the user equipment (UE) is blocked \cite{dardari2020communicating}. There is by now a large body of technical literature devoted to various communication-oriented applications \cite{8910627}, including reduced transmit power at active base stations for better energy efficiency, data rate coverage extension by illuminating dead zones, limited unintentional exposure to electromagnetic fields, enhanced privacy from generalized spatial filtering. 
More recently, it has become apparent that RISs are also beneficial to enrich the spatial awareness capabilities of future 6G systems, especially when combined with directive communications at millimeter wave (mmWave) frequencies. Therefore they can increase the amount of exploitable ``geometric"  deterministic location-dependent information conveyed by received multipath profiles and thereby, they boost positioning performances. An overview of the main opportunities, challenges and system candidates in the specific context of RIS-enabled localization and mapping is available in~\cite{wymeersch2019radio}. More specifically, several contributions have been focusing on exploiting the signal wavefront curvature at receiving RIS for direct low-complexity positioning~\cite{abu2020near,guidi2019radio,hu2018beyond}, whereas other works concern the use of RIS in reflection mode for 3D far-field localization and synchronization in single input single output (SISO), over successive multi-carrier (MC) downlink transmissions with a single RIS \cite{keykhosravi2020siso} or several RIS \cite{bjornson2021reconfigurable}. Finally, from a control standpoint, a joint bound-based RIS selection and phase profile optimization scheme has been proposed in~\cite{Wymeersch_ICC20}, which is shown to improve multipath-aided positioning accuracy (resp. positioning coverage) in comparison with multipath-aided positioning based on uncontrolled scattering (resp. uncontrolled reflections). All the aforementioned works on RIS-enabled positioning assume the LoS path to be present, whereas non-LoS (NLoS) propagation is seen as the dominating source of errors in conventional wireless localization systems, as well as a relevant operating context in most RIS-based communication use cases. To bridge this gap, a performance analysis of NLoS RIS-enabled localization is needed. 

\begin{figure}
    \centering
    \includegraphics[width=1\columnwidth]{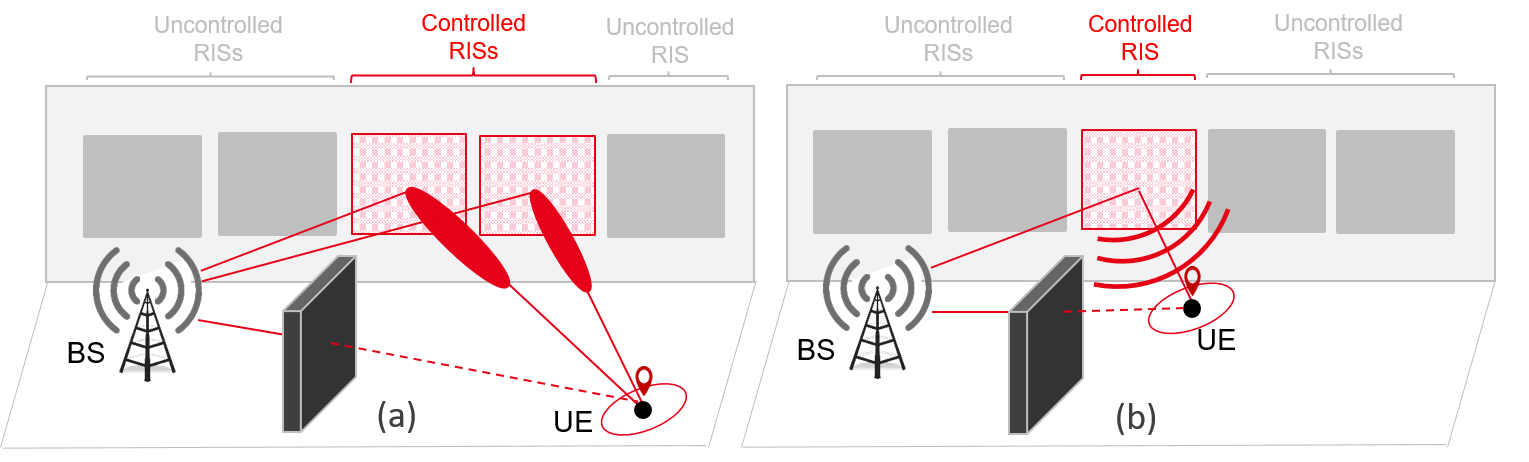}
    \caption{Obstructed line-of-sight in both far field (a) and near field (b) wireless localization with multiple RISs in reflection mode.}
    \label{fig:scenarios}
    \vspace{-6mm}
\end{figure}

In this paper, we aim at investigating more specifically the key issue of localization continuity under severe radio blockages, as shown in Fig.~\ref{fig:scenarios}. The main contribution of the paper is three-fold. Assuming SISO downlink positioning with RIS in reflection mode, we first propose a unified model and problem formulation to cover LoS and NLoS propagation situations, as well as both geometric near field (NF) and far field (FF) regimes. Next, we perform a Fisher information analysis characterizing the fundamental bounds on positioning performance, including identifiability considerations, as well as a sensitivity study with respect to key system parameters. Finally, we provide some insights regarding suitable system architecture and strategies that could support better localization continuity under severe radio blockage situations.


\subsubsection*{Notations} We denote the trace operator by $\text{tr}(\cdot)$ and the
Hadamard product by 
$\odot$, while $\text{diag}(\vect{x})$ returns a diagonal matrix with $\vect{x}$ on the diagonal. Vectors and matrices are indicated by lowercase and uppercase bold letters resp. The element in the $i$-th row and $j$-th column of matrix $\vect{A}$ is specified by $[\vect{A}]_{i,j}$. Similarly,
$[\vect{p}]_i$ indicates the $i$-th element of vector $\vect{p}$. The subindex
$i:j$ specifies all the elements between $i$ and $j$. All vectors
are columns, unless stated otherwise. The complex conjugate, Hermitian, and transpose operators are represented by $(\cdot)^*$, $(\cdot)^{\mathsf{H}}$, and $(\cdot)^\top$, respectively 

\section{Models and Problem Statement}\label{sec:model}

In this section, we describe the geometric model, as well as the signal and channel models. 
\subsection{Geometry Model}

We consider a single-antenna BS, a single-antenna UE, and {$K$ distinct} reflective planar RISs, each composed of $M$ elements, thus extending the SISO deployment scenario initially described in~\cite{keykhosravi2020siso}. The corresponding 3D locations, which are all assumed static, are expressed in the same global reference coordinates system: $\vect{p}_{\text{BS}} \in \mathbb{R}^{3 \times 1}$ is a vector containing the known BS coordinates, 
\textcolor{black}{$\vect{p}_{k} \in \mathbb{R}^{3 \times 1}$}
is a vector containing the known coordinates of the $k$-th RIS center, \textcolor{black}{$\vect{p}_{k,m} \in \mathbb{R}^{3\times 1}$} is a vector containing the known coordinates of the $m$-th element \textcolor{black}{belonging to the $k$-th RIS}, and $\vect{p}\in \mathbb{R}^{3\times 1}$ is a vector containing UE's unknown coordinates. The general problem is conceptually illustrated in Fig.~\ref{fig:scenarios}.

\subsection{Observation Model}
The BS broadcasts in downlink a wideband pilot signal $\vect{s}_t \in \mathbb{C}^{N\times 1}$ across a set of \textcolor{black}{$N$ subcarriers with frequency spacing $\Delta_f$, over $T$ successive} transmissions.
The complex signal vector $\vect{y}_t \in \mathbb{C}^{N\times 1}$ received by the UE at transmission $t$ is 
\begin{align}
\vect{y}_{t} =  \sum_{k=0}^{K} \beta_{k,t}    (\vect{s}_t \odot \vect{d}(\tau_{k})) + \vect{n}_t\label{eq:Received_signal}
\end{align}
where 
$\vect{n}_t \sim \mathcal{CN}(\vect{0},N_0 \vect{I}_{N})$ is the independent and identically distributed (i.i.d.) observation noise and $\vect{d}(\tau)=\left[1, \vect{e}^{-\jmath 2\pi\tau \Delta_{f}}, \ldots ,\vect{e}^{- \jmath 2\pi\tau (N-1)\Delta_{f}}
    \right]^{\top} \in  \mathbb{C}^{N\times 1}$, for 
\begin{align}
    \tau_{0} & =\norm{\vect{p}-\vect{p}_{\text{BS}}}/c + \Delta_{t},\\
    \tau_{k>0} & = \norm{\vect{p}-\vect{p}_{k}}/c + \norm{\vect{p}_{k}-\vect{p}_{\text{BS}}}/{c}
    + \Delta_{t},
\end{align}
in which 
$\Delta_{t}$ reflects the clock offset between the transmitter and the receiver 
and $\beta_{k,t}$ is the complex channel gain of the $k$-th path:
\begin{align}
    \beta_{0,t} & =\alpha_{0}, \forall t\\
    \beta_{k>0,t} & =\alpha_k \vect{a}^{\top}(\vect{p}_k,\vect{p})\text{diag}(\bm{\omega}_{k,t})\vect{a}(\vect{p}_{\text{BS}},\vect{p}_k)\\
    & = \textcolor{black}{\alpha_k}\vect{b}^{\top}(\vect{p}_k,\vect{p})\bm{\omega}_{k,t}.
\end{align}
Here, we have introduced time-invariant complex channel gain $\alpha_{0}$ for the LoS path and $\alpha_{k>0}$ for the paths via the $K$ RISs; 
$\bm{\omega}_{k,t}\in \mathbb{C}^{M\times 1}$ is the $t$-th phase profile vector applied to the $k$-th RIS over its $M$ elements, 
$\vect{b}(\vect{p}_k,\vect{p})=\vect{a}(\vect{p}_k,\vect{p})\odot\vect{a}(\vect{p}_{\text{BS}},\vect{p}_k)$. 
Note that, accordingly, the RIS is just assumed to be coarsely synchronized with the transmitter.
The steering vector $\vect{a}(\vect{p}_k,\vect{p}) \in \mathbb{C}^{N\times 1}$ is defined as having its $m$-th entry
\begin{align}
    [\vect{a}(\vect{p}_k,\vect{p})]_{m}=\exp\left(-\jmath\frac{2\pi}{\lambda}\left(\Vert\vect{p}-\vect{p}_{k,m}\Vert-d_k\right)\right),\label{eq:RIS_response_NF}
\end{align}
where $d_k=\Vert\vect{p}-\vect{p}_{k}\Vert$. When $d_k \gg \max_m \Vert \vect{p}_{k,m}-  \vect{p}_{k}\Vert $ (i.e., in far field), then 
\begin{align}\label{eq:RIS_response_FF}
[\vect{a}(\vect{p}_k,\vect{p})]_{m} \to \exp\left(-\jmath \vect{q}_{k,m}^\top \vect{k}(\psi_{\text{az},k},\psi_{\text{el},k}) \right),
\end{align}
where $\vect{q}_{k,m}=\vect{p}_{k,m}-\vect{p}_{k}$ and $\vect{k}(\psi_{\text{az},k},\psi_{\text{el},k})$ is the wavevector that depends on the angle of departure (AoD) at the $k$-th RIS respectively in elevation and azimuth, which is defined exactly like in~\cite{keykhosravi2020siso}, under the same orientation convention:
\begin{equation}
    \vect{k}(\psi_{\text{az},k},\psi_{\text{el},k})=-\frac{2 \pi}{\lambda}\left[\begin{array}{c}
\sin\psi_{\text{el},k} \cos \psi_{\text{az},k} \\
\sin\psi_{\text{el},k} \sin \psi_{\text{az},k}\\
\cos\psi_{\text{el},k}\\
\end{array}\right].
\end{equation}

Note that \eqref{eq:RIS_response_NF} is valid in both NF and FF, while the FF condition $d_k \gg \max_m \Vert \vect{p}_{k,m}-  \vect{p}_{k}\Vert$ depends on the RIS geometric size.

\section{Fisher Information Analysis}

The goal is to localize the UE from the observations $\vect{y}_1,\ldots, \vect{y}_T$ in both LoS and NLoS conditions (corresponding to $\alpha_{{0}} \neq 0$ and $\alpha_{{0}}=0$, respectively), depending on the NF or FF regime.

\subsection{General Approach}
The approach we follow comprises the following steps:
\begin{enumerate}
    \item Determination of the channel parameters in vector $\vect{\upzeta}_{\text{ch}}$ (e.g., angles, delays, gains...; See \eqref{eq:Estimation_param_NF} and \eqref{eq:Estimation_param_FF}) and computation of the FIM of the channel parameters \begin{align}
          \vect{\text{FIM}}_{\text{ch}}= \frac{2}{N_0} \sum_{t=1}^{T}  \Re\left\{\left(\frac{\partial\vect{\mu}_t} {\partial\vect{\upzeta}_{\text{ch}}}\right)^{\mathsf{H}} \frac{\partial\vect{\mu}_t} {\partial\vect{\upzeta}_{\text{ch}}}
    \right\},\label{eq:General_FIM}
    \end{align}
    where $\vect{\mu}_t$ denotes the noise-free observation at transmission $t$. 
    \item Determination of the position parameters $\vect{\upzeta}_{\text{po}}$ (position, clock bias, gains) with corresponding Jacobian $\vect{J}= {\partial\vect{\upzeta}_{\text{ch}}} /{ \partial\vect{\upzeta}_{\text{po}}}$ and computation of the FIM of the position parameters
    \begin{align}\label{eq:Positional_FIM}
    \vect{\text{FIM}}_{\text{po}}= \vect{J}^{\top}\vect{\text{FIM}}_{\text{ch}}\vect{J}.
\end{align}
\item Removal of the channel gains via the equivalent FIM to obtain the position and clock bias FIM, $\vect{\text{FIM}}_{\text{po}}(\vect{p},\Delta_t) \in \mathbb{R}^{4 \times 4}$. Finally, the position error bound (PEB, with unit meters), which characterizes a lower bound on the accuracy of any unbiased 3D location estimator, is  calculated as
\begin{align}
\vect{\text{PEB}} = \sqrt{\text{tr}\left(\left[\vect{\text{FIM}}_{\text{po}}^{-1}(\vect{p},\Delta_t)\right] _{(1:3,1:3)} \right)}.
\end{align}
Although our paper focuses mainly on position, it is also possible to derive a similar synchronization error bound (SEB, with unit seconds) in the same manner.
\end{enumerate}
 Under the assumptions that the RIS phase profiles among the different RISs are mutually orthogonal i.e., $\sum_t \bm{\omega}^{\mathsf{H}}_{k,t}\bm{\omega}_{k',t}=0$ when $k\neq k'$, each RIS provides independent information \cite{bjornson2021reconfigurable,keykhosravi2021semipassive}, which allows us to compute the Fisher information matrix (FIM) from each RIS separately and add up the information. In other words, we can compute a $4 \times 4$ FIM, say  $\vect{\text{FIM}}_{k,\text{po}}(\vect{p},\Delta_t)$, based \emph{only} on the $k$-th RIS (with associated channel parameters) and $\vect{\text{FIM}}_{0,\text{po}}(\vect{p},\Delta_t)$ based \emph{only} on the LoS path, with $\vect{\text{FIM}}_{\text{po}}(\vect{p},\Delta_t)=\sum_{k=0}^{K}\vect{\text{FIM}}_{k,\text{po}}(\vect{p},\Delta_t)$. This allows us to focus on a single RIS in the sequel (i.e., $K=1$), without loss of generality. 

\subsection{Specific Approach for NF}

The channel parameters are defined as 
\begin{align}\label{eq:Estimation_param_NF}
    \vect{\upzeta}_{\text{ch}} =[\vect{p}^{\top}, \tau_{1}, \tau_{0}, \alpha_{{1},{r}}, \alpha_{{1},{i}}, \alpha_{{0},{r}}, \alpha_{{0},{i}}]^{\top} \in \mathbb{R}^{9\times 1}
\end{align}
where $\alpha_k = \alpha_{{k},{r}}+ \jmath \alpha_{{k},{i}}$, $k \in \{0,1\}$. 
Accordingly, assuming the most generic RIS response (\ref{eq:RIS_response_NF}), the partial derivatives in~(\ref{eq:General_FIM}) are calculated as:
\begin{equation}
    \frac{\partial\vect{\mu}_{t}}{\partial\vect{p}}=
    \alpha_{{1}}
     (\vect{s}_t \odot \vect{d}(\tau_{1})) 
     \vect{a}^\top(\vect{p}_{\text{BS}},\vect{p}_{1})
     \text{diag}(\vect{\omega}_{t})
     \frac{\partial \vect{a}(\vect{p}_1,\vect{p})} {\partial\vect{p}} \label{eq:derivativePosition1}
\end{equation}
where $\frac{\partial \vect{a}(\vect{p}_1,\vect{p})} {\partial\vect{p}} \in \mathbb{C}^{M \times 3}$, with 
\begin{align}
    \frac{\partial\left[ \vect{a}(\vect{p}_1,\vect{p})\right]_m} {\partial\vect{p}} = 
    -\frac{\jmath 2\pi [\vect{a}(\vect{p}_1,\vect{p})]_m}{\lambda}\left(\vect{e}_{1,m}  - \vect{e}_1
    \right),\label{eq:derivatorPosition}
\end{align}
where $\vect{e}_{1,m}=({\vect{p} - \vect{p}_{1,m}})/{\norm{ \vect{p} - \vect{p}_{1,m} }}$ and $\vect{e}_1=({\vect{p} - \vect{p}_{1}})/{\norm{ \vect{p} - \vect{p}_{1} }}$ denote unit vectors pointing from the RIS to the UE location. Note that in FF $\vect{e}_{1,m}  - \vect{e}_1
     \to 0$, implying that in FF the position cannot be estimated from the path from 1 RIS. 
In addition, for $k\in \{0,1\}$
\begin{align}
    \frac{\partial\vect{\mu}_{t}}{\partial\tau_{k}}& =\beta_{k,t}\left(\vect{s}_t \odot \frac{\partial \vect{d}(\tau_{k})}{\partial\tau_{k}}
    \right), \label{eq:derivativeDelay}
\end{align}
where 
\begin{align}
    \frac{\partial \vect{d}(\tau_{k})}{\partial\tau_{k}} = -\jmath 2\pi\Delta_{f}\text{diag}([0,1,\ldots,N-1])\vect{d}(\tau_{k}).
\end{align}
Finally, the derivatives with respect to the channel gains are given by 
%
%
\begin{align}
    \left[\frac{\partial \vect{\mu}_{t}}{\partial\alpha_{{1},{r}}},\frac{\partial \vect{\mu}_{t}}{\partial\alpha_{{1},{i}} }\right]& =
     \vect{b}^{\top}(\vect{p}_1,\vect{p})\bm{\omega}_{1,t}(\vect{s}_t\odot \vect{d}(\tau_{1}))[1,\jmath ]\label{eq:derivativeGain1}\\
     \left[\frac{\partial \vect{\mu}_{t}}{\partial\alpha_{{0},{r}}},\frac{\partial \vect{\mu}_{t}}{\partial\alpha_{{0},{i}} }\right]& =(\vect{s}_t\odot \vect{d}(\tau_{0}))[1,\jmath]. \label{eq:derivativeGain2}
\end{align}
%
Substitution of \eqref{eq:derivativePosition1}--\eqref{eq:derivativeGain2} into \eqref{eq:General_FIM}, taking the real part, and summing over $T$ yields a numerical approach to compute 
the FIM of the channel parameters $\vect{\text{FIM}}_{\text{ch}}$. 

The position parameters are defined as 
\begin{align}
    \vect{\upzeta}_{\text{po}} =[\vect{p}^{\top}, \Delta_t, \alpha_{{1},{r}}, \alpha_{{1},{i}}, \alpha_{{0},{r}}, \alpha_{{0},{i}}]^{\top} \in \mathbb{R}^{8\times 1} 
\end{align}
with corresponding Jacobian $\vect{J} \in \mathbb{R}^{9 \times 8}$ given by
\begin{align}
    \vect{J} = \left[\begin{array}{ccc}
\vect{I}_{3} & \vect{0}_{3\times1} & \vect{0}_{3\times4}\\
 \vect{e}_{1}^{\top}/c & 1 & \vect{0}_{1\times4}\\
\vect{e}_{\text{BS}}^{\top}/c & 1 & \vect{0}_{1\times4}\\
\vect{0}_{4\times3} & \vect{0}_{4\times1} & \vect{I}_{4}
\end{array}\right], \label{eq:fullJacobian}
\end{align}
where $\vect{e}_{\text{BS}}=({\vect{p}-\vect{p}_{\text{BS}} })/{\norm{ \vect{p}-\vect{p}_{\text{BS}} }}$ and $\vect{e}_1$ was defined \eqref{eq:derivatorPosition}. Substitution of \eqref{eq:fullJacobian} into \eqref{eq:Positional_FIM} yields $\vect{\text{FIM}}_{\text{po}}$, from which the PEB and SEB are readily obtained. 

%
%
%
%

%

\subsection{Specific Approach for FF}
In far field, 
the vector of channel parameters is defined as
\begin{equation}
    \vect{\upzeta}_{\text{ch}}=[
    \tau_{1}, \tau_{0}, 
    \psi_{\text{az},1}, \psi_{\text{el},1}, \alpha_{1,{r}}, \alpha_{1,{i}}, \alpha_{{0},{r}}, \alpha_{{0},{i}}]^\top \in \mathbb{R}^{8\times 1}. \label{eq:Estimation_param_FF}
\end{equation}
Assuming (\ref{eq:RIS_response_FF}) for the RIS response, the partial derivatives in (\ref{eq:General_FIM}), as well as the resulting positional FIM for each single RIS contribution independently can be calculated similar to \cite{keykhosravi2020siso} and the extension to $K$ RIS is straightforward. 
\subsection{Identifiability Conditions} 
Based on FIM analysis, the following identifiablity conditions hold in the wideband regime (i.e., $N\gg 1$):
\begin{itemize}
    \item \emph{Only the LoS path is present ($K=0$): } the UE cannot be localized, neither in NF, nor FF. Only when 4 BS are present, 3D localization and synchronization becomes possible. 
    \item \emph{LoS and NLoS paths are both present ($K>0$, $\alpha_0\neq 0$):} in NF and FF, the UE can be localized, provided a sufficiently large number of transmissions is used.\footnote{For example, estimation of the AoD from 1 RIS requires at least $T=2$ transmissions with non-parallel $\vect{\omega}_t$ \cite{keykhosravi2020siso}. Estimation of the position in NF requires at least $T=3$ transmissions \cite{abu2020near}. }  
    \item \emph{Only NLoS paths are present ($K>0$, $\alpha_0=0$):} the UE can be localized with a single RIS in NF, but not in FF. However, with $K\ge 2$ RIS, the UE can also be localized in FF (by the intersection of $K$ lines in 3D, based on the AoD estimates). 
\end{itemize}

%
%

%
\section{Analysis of Theoretical Positioning Performances}\label{sec:fim_analysis}
\subsection{Simulation Parameters and Settings}\label{subsec:Simulation_settings}
%

\begin{figure}
    \centering
    \includegraphics[width=1\columnwidth]{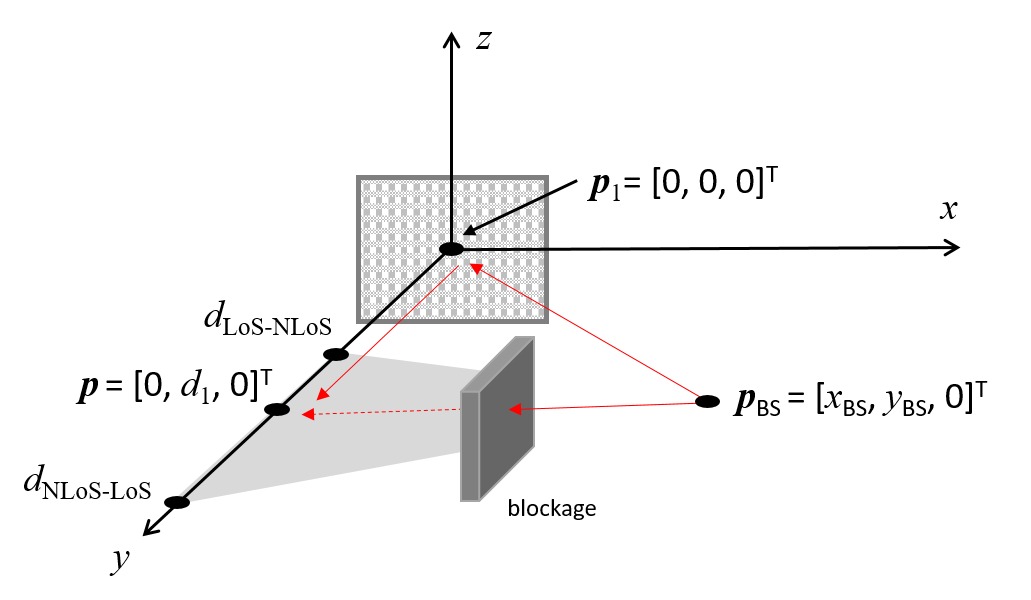}
    \vspace{-10mm}
    \caption{Single-RIS scenario considered for PEB evaluations. The represented blocking obstacle is considered mainly in the heatmap representation of Fig.~\ref{fig:Place_holder_heatmap}, whereas the particular UE location shown along the $y$ axis holds mainly in Figs.~\ref{fig:Place_Holder_1D_PEB_Single_RIS_Varying_Size}--\ref{fig:Place_Holder_1D_PEB_NLoS} (with $d_1 =\norm{\vect{p}-\vect{p}_1}$).} 
    \label{fig:Simulation_scenario_parametric_evaluation}
    \vspace{-4mm}
\end{figure}

Based on the PEB characterized in the previous section, we hereafter analyze the theoretical positioning performance in the particular case when $K=1$, as a function of operating conditions and main system parameters, by means of numerical evaluations. Without loss of generality, we assume the transmission of $T=25$ consecutive symbols $\vect{s}_t =\sqrt{E_s}\vect{\text{I}}_{N\times 1}$, $\forall t$ over $N=3000$ subcarriers with subcarrier spacing $\Delta_f=120$ kHz at the center frequency of $28$ GHz (or equivalently, with an average wavelength $\lambda=1$ cm). 
The total transmission power (i.e., $N \Delta_f E_s$) is set to 20 dBm and the UE noise figure to $8$ dB, besides typical noise power spectral density $N_0=-174$ dBm/Hz. We also consider $M \in \{32\times 32, 64\times 64, 128\times 128\}$ RIS elements with an inter-element distance of $\lambda/2$ and random phase profiles\footnote{RIS phase profiles are balanced (i.e., $\sum_t \vect{\omega}_t \approx \vect{0}$) for large $M$, improving multipath resolution. However, without loss of generality, other optimized profiles could have been chosen  ~\cite{Wymeersch_ICC20,abu2020near}). } $\vect{\omega}_{1,t}$, $\forall t$.   \textcolor{black}{The amplitude of the channel gains $\alpha_k$, $k \in \{0,1\}$ are  calculated based on Friis’ formula and their phases are set randomly in $[0, 2\pi)$.}
In terms of explored scenarios (See Fig.~\ref{fig:Simulation_scenario_parametric_evaluation}), we assume the RIS to be placed on the $xz$ plane (i.e., perpendicular to the $y$ axis) and centered in $\vect{p}_1=[0,0,0]^{\top}$, one single BS  $\vect{p_{\text{BS}}}=[x_{\text{BS}},y_{\text{BS}},0]^{\top}$ (set to $\vect{p_{\text{BS}}}=[5,5,0]^{\top}$ unless otherwise stated), and a single UE, which can occupy either any location in the scene. 


%

%
%

\subsection{Simulation Results and Discussions}
\begin{figure}
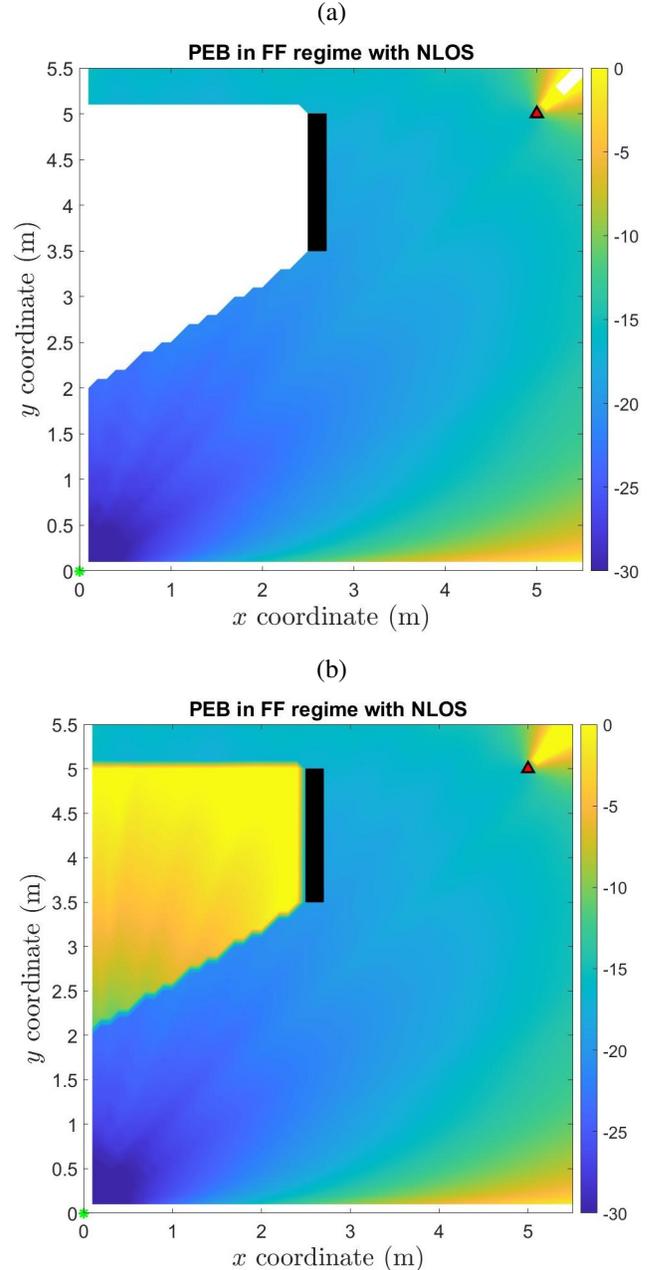

    \centering
    (a)
    \resizebox{1\columnwidth}{!}{\input{Figures/FF}}\\
    (b)
    \resizebox{0.975\columnwidth}{!}{\input{Figures/NF}}
    \vspace{-6mm}
    \caption{PEB heatmap (in dB) for $z=0$, as a function of UE location for far field (a) and near field (b) models, with 1 single RIS with $M=32\times 32$ in $[0,0,0]^\top$m, 1 BS in $[5,5,0]^\top$m and a finite obstacle with boundaries $[2.5,3.5,0]^\top$m and $[2.5,5,0]^\top$m (i.e., parallel to the $y$ axis). Note that the far field PEB in the NLoS region is infinite (localization is not feasible).}
    \label{fig:Place_holder_heatmap}
    \vspace{-4mm}
\end{figure}

On Fig.~\ref{fig:Place_holder_heatmap}, we first show the PEB heatmap (in dB scale, where $0$ dB corresponds to $1$ meter, $-10$ dB to $0.1$ meter, etc) with {$M=32 \times 32$} in both FF (a) and NF (b) regimes, as a function of UE location in a room of {$5.5~\text{m} \times  5.5~\text{m}$,}, conditioned upon LoS/NLoS with respect to the BS in $[5,5,0]^\top$m, while assuming a finite obstacle delimited by boundaries in $[2.5,3.5,0]^\top$m and $[2.5,5,0]^\top$m (i.e., parallel to the $y$ axis). 
We can make a number of observations: the NF PEB is always smaller than the FF PEB, due to the exploitation of wavefront curvature. In both NF and FF, positioning performance is better closer to the RIS, since the positioning is fundamentally limited by the weaker RIS path. In FF, there are two regions where the PEB in FF is infinite (i.e., the FIM is singular): the shadowed region where the LoS path is unavailable, and the locations behind the BS, where the FF ToA information is not informative \cite{bjornson2021reconfigurable}. In contrast, the NF PEB is finite for all locations and the exploitation of one single RIS-reflected path makes localization feasible, even if accuracy is significantly degraded (typically, by approximately one order of magnitude, from decimeter to meter levels). 

%

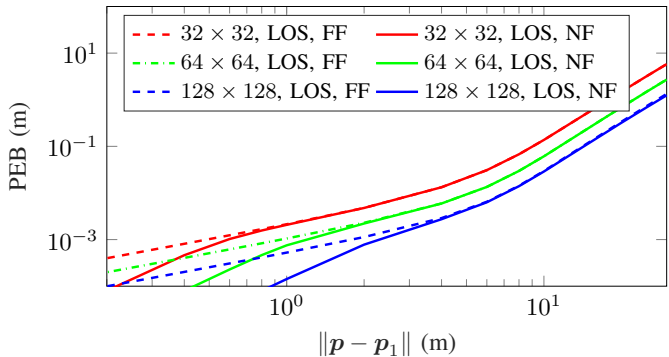
\begin{figure}
    \centering
    \resizebox{1\columnwidth}{!}{
%
%
\begin{tikzpicture}

\begin{axis}[%
width=8cm,
height=4cm,
at={(1.011in,0.642in)},
scale only axis,
xmode=log,
xmin=0.2,
xmax=30,
xminorticks=true,
xlabel style={font=\color{white!15!black}},
xlabel={$\norm{\vect{p}-\vect{p}_1}$ (m)},
ymode=log,
ymin=1e-04,
ymax=100,
yminorticks=true,
ylabel style={font=\color{white!15!black}},
ylabel={PEB (m)},
axis background/.style={fill=white},
legend style={legend cell align=left, align=left, draw=white!15!black},
legend pos=north west,
legend columns=2,
]
\addplot [color=red, dashed,line width=1.0pt]
  table[row sep=crcr]{%
0.01	3.27409677825949e-05\\
0.02	4.50580327030239e-05\\
0.03	6.14402294190084e-05\\
0.04	7.94524122532985e-05\\
0.05	9.82480124381133e-05\\
0.06	0.000117458140656199\\
0.07	0.000136946300987036\\
0.1	0.000196335302248646\\
0.2	0.000398291673483173\\
0.4	0.000811701477729711\\
0.6	0.00123728254081438\\
0.8	0.00167745359060928\\
1	0.00213465373370175\\
2	0.00477755087140558\\
4	0.0132705690079579\\
6	0.0308533062168294\\
8	0.0676462036841984\\
10	0.137437747577923\\
15	0.551838527353413\\
20	1.48726429851195\\
25	3.16429670937508\\
30	5.80316458590026\\
};
\addlegendentry{\small{$32\times 32$, LOS, FF}}

\addplot [color=red,line width=1.0pt]
  table[row sep=crcr]{%
0.01	1.9230779471565e-06\\
0.02	2.14669104349471e-06\\
0.03	2.80694472614063e-06\\
0.04	3.72860363654371e-06\\
0.05	4.90799694559718e-06\\
0.06	6.37231822002918e-06\\
0.07	8.15893711460361e-06\\
0.08	1.03085783812187e-05\\
0.0900000000000001	1.28634606227303e-05\\
0.1	1.58680063046741e-05\\
0.2	8.01026590977846e-05\\
0.4	0.000462203457316919\\
0.6	0.0010283677635784\\
0.8	0.0015660665961521\\
1	0.00207154912593525\\
2	0.00476770126476621\\
4	0.0132683553288951\\
6	0.0308508621113893\\
8	0.0676404271496689\\
10	0.137421820162235\\
15	0.551731484891036\\
20	1.48688711693396\\
25	3.16326659391883\\
30	5.80095638890762\\
};
\addlegendentry{\small{$32\times 32$,  LOS, NF}}

\addplot [color=green, dashdotted,line width=1.0pt]
  table[row sep=crcr]{%
0.01	1.63764564874327e-05\\
0.02	2.25291943493055e-05\\
0.03	3.07231050731667e-05\\
0.04	3.97246113194047e-05\\
0.05	4.91146356217026e-05\\
0.06	5.87267361604377e-05\\
0.08	7.83071907684645e-05\\
0.1	9.8152216183062e-05\\
0.2	0.000199039197143627\\
0.4	0.000404969799528277\\
0.6	0.000615767630632646\\
0.8	0.000831945198119025\\
1	0.00105426532762311\\
2	0.00229038341565355\\
4	0.00596180886035051\\
6	0.0134874957330394\\
8	0.0297277279875041\\
10	0.0612390152950383\\
20	0.68811629376417\\
25	1.4741429182166\\
30	2.71477683856173\\
};
\addlegendentry{\small{$64\times 64$, LOS, FF}}

\addplot [color=green,line width=1.0pt]
  table[row sep=crcr]{%
0.01	1.41003597279184e-06\\
0.02	1.35422045166687e-06\\
0.03	1.55897590962835e-06\\
0.04	1.84851696367033e-06\\
0.05	2.19415213095934e-06\\
0.06	2.59064897751313e-06\\
0.07	3.03895364804904e-06\\
0.08	3.54216124825886e-06\\
0.0900000000000001	4.10439166555524e-06\\
0.1	4.73014652331932e-06\\
0.2	1.56247835583975e-05\\
0.4	7.95998504285132e-05\\
0.6	0.000228362618797284\\
0.8	0.000464621292787929\\
1	0.000752700614245324\\
2	0.00221202261968248\\
4	0.00594079854989451\\
6	0.0134660635531969\\
8	0.0296845931449425\\
10	0.0611360127815578\\
20	0.685731537923434\\
25	1.4679414605845\\
30	2.70201288762491\\
};
\addlegendentry{\small{$64\times 64$, LOS, NF}}

\addplot [color=blue, dashed, line width=1.0pt]
  table[row sep=crcr]{%
0.01	8.18756029376008e-06\\
0.02	1.12644496961097e-05\\
0.03	1.53611334021512e-05\\
0.04	1.98615306764648e-05\\
0.05	2.45577229103111e-05\\
0.06	2.93638955176812e-05\\
0.07	3.42357393375074e-05\\
0.08	3.91529822424895e-05\\
0.0900000000000001	4.41022983216701e-05\\
0.1	4.90757690468643e-05\\
0.2	9.9505517901522e-05\\
0.4	0.000202379465725721\\
0.6	0.000307523021812498\\
0.8	0.000415127538373768\\
1	0.000525492155628943\\
2	0.00113257370268203\\
4	0.0028907095210764\\
6	0.00647897196558606\\
8	0.0143068183964152\\
10	0.0296117189095345\\
15	0.122933837560334\\
20	0.336766562103505\\
25	0.722974393379541\\
30	1.33327863137298\\
};
\addlegendentry{\small{$128\times128$, LOS, FF}}

\addplot [color=blue,line width=1.0pt]
  table[row sep=crcr]{%
0.01	1.13422093322343e-06\\
0.02	9.95684662364145e-07\\
0.03	1.0602020901192e-06\\
0.04	1.17147565113766e-06\\
0.05	1.3027549640803e-06\\
0.06	1.44698802101022e-06\\
0.07	1.60204226596769e-06\\
0.08	1.76732978295935e-06\\
0.0900000000000001	1.94287333879446e-06\\
0.1	2.12899599295365e-06\\
0.2	4.68561627426677e-06\\
0.4	1.55652006377956e-05\\
0.6	3.87265263577547e-05\\
0.8	7.95295248270777e-05\\
1	0.000142266975019664\\
2	0.00077893331594684\\
4	0.00273460180425875\\
6	0.0063159062578735\\
8	0.0139879423713085\\
10	0.0288655515753207\\
15	0.118289242515442\\
20	0.320046754221022\\
25	0.680178244349008\\
30	1.2443997307649\\
};
\addlegendentry{\small{$128\times 128$, LOS, NF}}

\end{axis}
\end{tikzpicture}%
}
    \vspace{-8mm}
    \caption{LoS PEB in both NF and FF regimes, as a function of the distance $\norm{\vect{p}-\vect{p}_1}$ between a UE in $[0,\norm{\vect{p}-\vect{p}_1},0]^\top$m and $K=1$ RIS of size $M \in \{32\times 32,64\times 64,128\times 128\}$ in $[0,0,0]^\top$m, with 1 BS in $[5,5,0]^\top$m.} 
    \label{fig:Place_Holder_1D_PEB_Single_RIS_Varying_Size}
\end{figure}

Fig.~\ref{fig:Place_Holder_1D_PEB_Single_RIS_Varying_Size} shows the PEB in LoS condition, as a function of the distance between the UE and the RIS, $\norm{\vect{p}-\vect{p}_1}$, as if the UE was following a 1D trajectory along the $y$ axis from the RIS, with  $M \in \{32\times 32,64\times 64,128\times 128\}$.
Here again, the NF PEB is better than the FF PEB and the relative gain is especially significant at shorter distances to the RIS. The reference distance at which the NF PEB has converged to the FF PEB, as well as the final gap between the two PEBs after convergence, both depend on the RIS size: for the smaller RIS ($M = 32 \times 32$), NF and FF PEB coincide after about 4 meters, while for the larger RIS, a gap is visible even beyond 20 meters. 
Obviously, larger $M$ values also globally improve performance in both FF and NF regimes. In NF for instance, each time the number of elements per dimension is doubled, achievable accuracy improved with about 
one order of magnitude. 

\begin{figure}
    \centering
    \resizebox{1\columnwidth}{!}{
%
%
\definecolor{mycolor1}{rgb}{0.00000,0.44700,0.74100}%
\begin{tikzpicture}

\begin{axis}[%
width=8cm,
height=4cm,
at={(1.011in,0.642in)},
scale only axis,
xmode=log,
xmin=0.2,
xmax=30,
xminorticks=true,
xlabel style={font=\color{white!15!black}},
xlabel={$\norm{\vect{p}-\vect{p}_1}$ (m)},
ymode=log,
ymin=1e-4,
ymax=100,
yminorticks=true,
ylabel style={font=\color{white!15!black}},
ylabel={PEB (m)},
axis background/.style={fill=white},
legend style={at={(0.03,0.97)}, anchor=north west, legend cell align=left, align=left, draw=white!15!black}
]
\addplot [color=mycolor1,line width=1.0pt]
  table[row sep=crcr]{%
0.01	1.9230779471565e-06\\
0.02	2.14669104349471e-06\\
0.03	2.80694472614063e-06\\
0.04	3.72860363654371e-06\\
0.05	4.90799694559718e-06\\
0.06	6.37231822002918e-06\\
0.07	8.15893711460361e-06\\
0.08	1.03085783812187e-05\\
0.0900000000000001	1.28634606227303e-05\\
0.1	1.58680063046741e-05\\
0.2	8.01026590977846e-05\\
0.4	0.000462203457316919\\
0.6	0.0010283677635784\\
0.8	0.0015660665961521\\
1	0.00207154912593525\\
2	0.00476770126476621\\
4	0.0132683553288951\\
6	0.0308508621113893\\
8	0.0676404271496689\\
10	0.137421820162235\\
15	0.551731484891036\\
20	1.48688711693396\\
25	3.16326659391883\\
30	5.80095638890762\\
};
\addlegendentry{\small{$32\times 32$, LOS, NF}}

\addplot [color=red,line width=1.0pt]
  table[row sep=crcr]{%
0.01	1.92462135579134e-06\\
0.02	2.14800973105558e-06\\
0.03	2.80876885618469e-06\\
0.04	3.7311604803017e-06\\
0.05	4.91201047220193e-06\\
0.06	6.37901172015407e-06\\
0.07	8.16952965952151e-06\\
0.08	1.03252289327342e-05\\
0.0900000000000001	1.28901743445148e-05\\
0.1	1.59099623818101e-05\\
0.2	8.16357213374985e-05\\
0.4	0.000559088049137689\\
0.6	0.00182721710175131\\
0.8	0.00428143358572522\\
1	0.00831712400519524\\
2	0.0660558142694641\\
10	8.23771018512028\\
30	222.398903575885\\
};
\addlegendentry{\small{$32\times 32$, NLOS, NF}}

\addplot [color=red, dashdotted,line width=1.0pt]
  table[row sep=crcr]{%
0.01	1.41256504766874e-06\\
0.02	1.35570222453761e-06\\
0.03	1.56028587302255e-06\\
0.04	1.84985733909999e-06\\
0.05	2.19563674424355e-06\\
0.06	2.59237813165441e-06\\
0.07	3.04102720821577e-06\\
0.08	3.54471612545916e-06\\
0.0900000000000001	4.10753929597977e-06\\
0.1	4.73409939223749e-06\\
0.2	1.56642449557332e-05\\
0.4	8.10575664476213e-05\\
0.6	0.000245127836577353\\
0.8	0.000557166600392195\\
1	0.00106651333207967\\
2	0.00829958968745036\\
4	0.0659301318886838\\
15	3.4691589162079\\
30	27.7497679600484\\
};
\addlegendentry{\small{$64\times 64$, NLOS, NF}}

\addplot [color=red, dashed]
  table[row sep=crcr]{%
0.01	1.13962115351159e-06\\
0.02	9.98167704653774e-07\\
0.03	1.06197143192218e-06\\
0.04	1.17295846915006e-06\\
0.05	1.30410180837754e-06\\
0.06	1.44828808561618e-06\\
0.07	1.60333448457182e-06\\
0.08	1.76865052229306e-06\\
0.0900000000000001	1.94425936817404e-06\\
0.1	2.13045399887689e-06\\
0.2	4.68946392214746e-06\\
0.4	1.56032004302303e-05\\
0.6	3.90014660194516e-05\\
0.8	8.09135926832265e-05\\
1	0.000147476584314566\\
2	0.00106574271756823\\
4	0.00829521776121828\\
6	0.0278517143628263\\
8	0.0658987957516204\\
10	0.128600000686276\\
15	0.433662994541285\\
20	1.02764154795162\\
25	2.00684085177862\\
30	3.46756609503463\\
};
\addlegendentry{\small{$128\times 128$, NLOS, NF}}

\end{axis}

\end{tikzpicture}%
}
    \vspace{-8mm}
    \caption{PEB in NF regime, as a function of the distance 
    $\norm{\vect{p}-\vect{p}_1}$   
    between a UE in $[0,\norm{\vect{p}-\vect{p}_1},0]^\top$m and $K=1$ RIS with $M=32\times 32$ and $M=64\times 64$ elements in $[0,0,0]^\top$m, with 1 BS in $[5,5,0]^\top$m.}
    \label{fig:Place_Holder_1D_PEB_NLoS}
    \vspace{-4mm}
\end{figure}
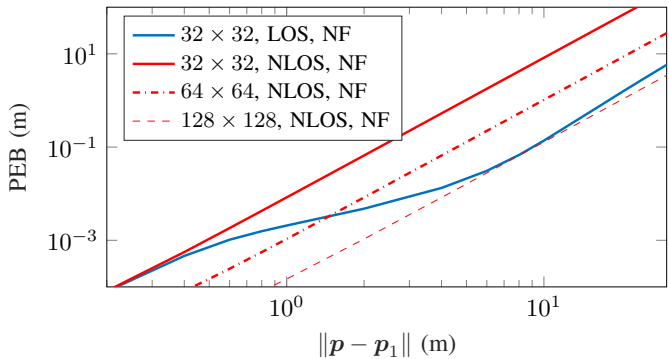

 Finally, Fig.~\ref{fig:Place_Holder_1D_PEB_NLoS} shows a similar PEB representation as a function of $\norm{\vect{p}-\vect{p}_1}$ with $M \in \{32\times 32,64\times 64\}$ in the NF regime only, while comparing both LoS and NLoS conditions. 
 In NLoS, it is noticed again that NF makes coarse single-RIS localization feasible, contrarily to FF. Moreover, the NF PEB with $M=64 \times 64$ in NLoS outperforms the NF PEB with $M=32 \times 32$ in LoS. This confirms that large RIS sizes (either physically large in static hardware settings and/or electronically expandable on-demand through dynamic control) could compensate the temporary loss of the direct path information to some extent.  
%


%
%

Overall, the various observations above tend to suggest that when a LoS-to-NLoS transitions is detected at system level along the UE trajectory (e.g., through innovation monitoring in standard Kalman tracking filters), exploiting one single RIS reflection in NF could maintain localization capabilities (even if degraded), and hence, could preserve service continuity, given that the RIS size is sufficient large and that high signal processing complexity is affordable on the receiver side to interpret the signal wavefront curvature for direct positioning. In FF on the contrary, multi-RIS operations are likely required to restore non-ambiguous localization capabilities in NLoS, thus shifting system complexity onto data association and RIS selection problems. 
\section{Conclusion}
In this paper, we have characterized and analyzed the theoretical positioning performance of SISO MC downlink multipath-aided localization in both LoS and NLoS conditions, while assuming one RIS in reflection mode. 
Numerical PEB evaluations in a canonical scenario confirm that, whenever the UE is close enough to the RIS and/or when the RIS is large, exploiting the signal wavefront curvature of the RIS-reflected multipath component in NF could be sufficient to directly infer user’s position in the absence of direct path, even if the achievable NLoS accuracy is shown to be relatively low with the chosen system parameters. 
Seamless and automated NLoS mitigation strategies could be proposed at system-level to contextually chose the number of controlled elements per RIS (in NF especially) or to activate multi-RIS processing (in FF especially) only if needed, hence minimizing complexity accordingly (typically, based on the latest estimated UE location or some prior knowledge).
Other future works concern the derivation of closed-form positional FIM expressions in FF and NF, the injection of prior information about UE's location and uncertainty in the latter FIM for continuous RIS optimization and localization refinement, the design of practical estimation algorithms to exploit the NF localization capabilities, as well as multi-user RIS-enabled localization schemes in a shared physical environment.  
\section*{Acknowledgement}
This work has been supported by H2020 RISE-6G project.

\bibliographystyle{IEEEtran}
\bibliography{references}

\end{document}